\documentclass[12pt,english]{article}

\usepackage[T1]{fontenc}
\usepackage[latin9]{inputenc}
\usepackage[a4paper]{geometry}
\usepackage{color}
\definecolor{document_fontcolor}{rgb}{0, 0, 0}
\color{document_fontcolor}
\usepackage{amstext}
\usepackage{graphicx}
\usepackage{setspace}
\usepackage{amsmath}
\usepackage{mathtools}
\usepackage{amssymb}
\makeatletter

\providecommand{\tabularnewline}{\\}

\newenvironment{lyxlist}[1]
{\begin{list}{}
{\settowidth{\labelwidth}{#1}
 \setlength{\leftmargin}{\labelwidth}
 \addtolength{\leftmargin}{\labelsep}
 }}
{\end{list}}

\@ifundefined{date}{}{\date{}}

\@ifundefined{definecolor}
 {\usepackage{color}}{}
\usepackage{array}\usepackage{amsthm}\usepackage{float} \makeatletter

\newcommand{\lyxmathsym}[1]{\ifmmode\begingroup\def\b@ld{bold}
  \text{\ifx\math@version\b@ld\bfseries\fi#1}\endgroup\else#1\fi}
\providecommand{\tabularnewline}{\\}

 \@ifundefined{definecolor}{

}{\makeatletter} \AtBeginDocument{
  
}

\providecommand{\tabularnewline}{\\}

\usepackage{fancyhdr}\usepackage{titlesec}\titleformat{\chapter}[display]
  {\normalfont\huge\bfseries\centering}{\MakeUppercase\chaptertitlename\ \thechapter}{20pt}{\LARGE}

\usepackage[square, numbers, comma, sort&compress]{natbib}

\makeatother

\usepackage{babel}
\begin{document}

\title{FSO-QKD protocols under free-space losses and device 
imperfections: a comparative study}

\author{Mitali Sisodia\thanks{\emph{Email: mitalisisodiyadc@gmail.com}},
Omshankar, Vivek Venkataraman and Joyee Ghosh\thanks{\emph{Email: joyee@physics.iitd.ac.in}}}
\maketitle
\begin{center}
Quantum Photonics Lab, Department of Physics, Indian Institute of Technology Delhi, New Delhi, 110016, India
\par\end{center}

\begin{abstract}
Quantum key distribution (QKD) is a technique to establish a secret key between two parties through a quantum channel. Several QKD protocols have been proposed and implemented over optical fibers or free-space links. The main challenge of operating QKD protocols over a free-space link is atmospheric losses. 
In this paper, we have studied and compared the performance of single and entangled photon based QKD protocols by evaluating the quantum bit error rate and secure key rate for terrestrial free-space quantum communication by considering different free-space losses, such as geometrical losses, atmospheric losses as well as device imperfections. 
\end{abstract}
\begin{lyxlist}{00.00.0000}
\item [{\textbf{Keywords:}}] Quantum key distribution, Quantum bit error
rate, Secure key rate, Geometrical losses, Atmospheric losses, Bell
parameter.
\end{lyxlist}

\section{Introduction}
Classically, the security of  information sent from one place to the other is mainly based on the popular Rivest-Shamir-Adleman (RSA) algorithm  \cite{time_pad}, which relies on the computationally extensive task of factorizing the product of two large prime numbers. This method hinges on the computational complications of certain mathematical tasks and is thus vulnerable to technological progress. Hence, there is a need to develop robust protocols to secure  the shared information. Quantum cryptography and in particular, quantum key distribution (QKD) fulfills this criterion by taking  advantage of the fundamental quantum physical properties such as (i) no cloning theorem, where any unknown quantum state cannot be copied;
(ii) measurement collapse the quantum system to one of the possible quantum state, and (iii) irreversiblity of measurements implying that an output state cannot be used to generate an input state. 
The first QKD protocol was proposed by C. H. Bennett and G. Brassard in 1984 \cite{BB84} utilizing qubits encoded through the polarization property of single photons. Later, several QKD protocols and
their security proofs against an eavesdropper have been studied \cite{six-state,Ekert,bbm_92,QKD2,QKD6,EPR,Gisin,six-state1,six-state2,koashi,Rev2020,BB84_security}.
Various sources of single and/or entangled photons (flying qubits) have proved to be good candidates for quantum communication.


In 1989, the first table-top QKD experiment has been performed over a 
32 cm quantum channel length \cite{32cm}. Quantum communication schemes can be performed using optical fibers, terrestrial free-space optical (FSO), and satellite-based FSO implementations. Although there have been numerous implementations of QKD protocols that are optical-fiber based, the length of communication achieved have been  only a few hundreds of kilometeres due to the limitation of an exponential increase of fiber losses with length \cite{OF1,of2,of3,OF_mdi}. On the other hand, a FSO (terrestrial and satellite) channel has proved to be a promising quantum channel providing secure quantum communication for longer distances (globally) and overcoming the problem of limited distant quantum communication through fibers \cite{FSO1,FSO2,FSO5,Inter_2_QBER,inter6,inter7_Qber,inter8,BBM1,Daylight,BBM3,BBM4}
for longer distances. The main challenge of implementing QKD protocols over free-space link is due to atmospheric losses \cite{atmoshperic_effects}. Other parameters such as timing,
weather, protocol, place (ground), etc. \cite{weather} also play important roles. The free-space losses can be broadly categorized into  geometrical losses and atmospheric losses. Every protocol has distance limitations due to these losses that grow with the transmission distance. Several studies mention these losses however a thorough quantitative study of the effect of the
free-space losses (particularly geometrical and atmospheric losses)
for various standard QKD protocols are not sufficiently investigated so far. Some of the FSO-based QKD protocols are- Bennett and Brassard, 1984 (BB84) \cite{BB84}, six-state \cite{six-state}, Ekert, 1991 (E91) \cite{Ekert}, and Bennett, Brassard, and Mermin, 1992 (BBM92) \cite{bbm_92}. Their performance comparison for particular
distances (different length scales) have not been studied in detail.
In this paper, we have theoretically compared the performance of free-space prepare-and-measure-based
(BB84 and six-state) and entanglement-based (BBM92 and Ekert91) QKD
protocols for different channel lengths and studied the effect of atmospheric losses on the quantum bit error rate (QBER) and the secure key rate (SKR).  We have also explored the impact of device imperfections through detection efficiencies,  losses and other parameters.

The paper is organized as follows: in Sections \ref{sec:Quantum-Bit-Error} and \ref{sec:Free-Space-losses}, we discuss QBER, SKR and free-space losses, the calculation of QBER and SKR for single-photon based QKD protocols (BB84 and six-state) are discussed in Section \ref{sec: BB84_and_B92}. In Section \ref{sec:Ekert91-and-BBM92},
entangled-photon based QKD protocols (E91 and BBM92) have been discussed. Finally, in Section \ref{sec:Conclusion}, we have concluded the results. 
\begin{figure}
\begin{centering}
\includegraphics[scale=0.6]{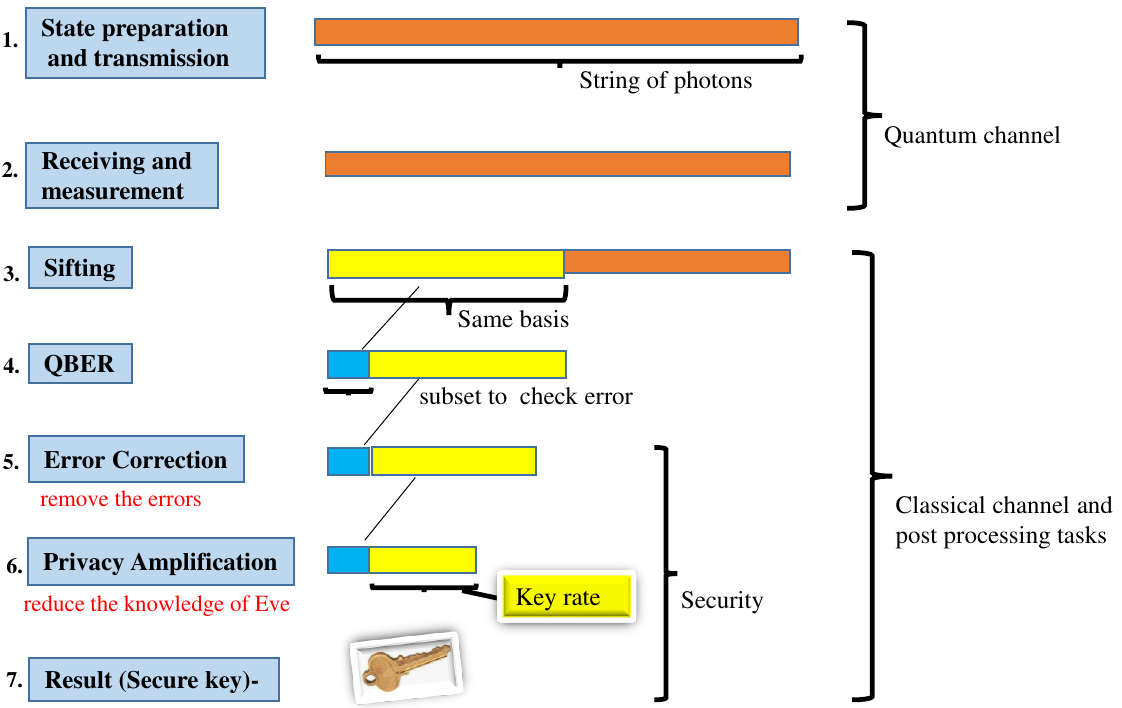}
\par\end{centering}
\caption{\label{fig:Steps-for-performing}Flow chart showing the steps involved in a general QKD protocol.
In Ekert's protocol, one step is added that is Bell's inequality test which is performed during sifting process. }
\end{figure}

\section{Quantum Bit Error Rate (QBER) and Secret Key rate (SKR)\label{sec:Quantum-Bit-Error}}
Quantum bit error rate (QBER) evaluates the information leakage to an unauthorized
third party (Eve) or due to imperfection of the physical devices. It is an important criterion to evaluate the performance of various QKD systems and has been calculated for various fiber-based or free-space QKD protocols \cite{OF1,of2,of3,OF_mdi,FSO1,FSO2,FSO5,
Inter_2_QBER,inter6,inter7_Qber,inter8,BBM1,Daylight}. 
It is defined as the ratio of the wrong bits to the
total number of bits received and can be expressed as \cite{Gisin} 
\begin{equation}
Q=\frac{N_{wrong}}{N_{total}}=\frac{N_{error}}{N_{correct+error}}\cong\frac{N_{error}}{N_{sift}},\label{eq:1}
\end{equation}
where $N_{sift}$ is the sifted key when Alice and Bob choose the compatible basis and $N_{error}$(<<$N_{sift}$) represents the error in the total number of bits.
When, the QBER that determines the security of the QKD protocol (ideally $Q = 0$; lower the value of QBER, higher the security of the protocol or vice versa) is higher than the threshold value (vary for different QKD protocols; e.g., 11\% for BB84 and 12.6\% for six-state protocol \cite{Rev2020}) then they discard and repeat the QKD protocol. The non-zero value of QBER is due to free-space losses, Eve\textquotesingle s  
presence, noises, imperfections in the physical devices, etc. A secret key can only be generated if the  mutual information $\left(I_{AB}\right)$ of Alice and Bob is greater than Eve's information $\left(I_{E(n)}\right)$  \cite{QKD2,Rev2020}. The secret key rate (SKR) is expressed as 
\begin{equation}
\text{\ensuremath{S=\text{\ensuremath{n_{sift}\,\left[I_{AB}-I_{E(n)}\right],}}}}\label{eq:skr}
\end{equation}
where $I_{AB}=1-h(Q)\,=\,1+Q\,\text{lo\ensuremath{g_{2}}}(Q)+(1-Q)\,\text{lo\ensuremath{g_{2}}}(1-Q)$, $h(Q)$ is the Shannon entropy, $h(Q)=-Q\,\text{lo\ensuremath{g_{2}}}(Q)-(1-Q)\,\text{lo\ensuremath{g_{2}}}(1-Q)$ and $n_{sift}$ is the basis reconciliation factor
(when Alice and Bob choose compatible bases).

\section{Free-space losses (Geometrical/Atmospheric) \label{sec:Free-Space-losses}}
The main challenge of implementing QKD protocols over free-space optical (FSO) channel is the free-space losses such as absorption, scattering, diffraction,
turbulence, etc. and are categorized in two parts (1) geometric (2) atmospheric \cite{QKD2} that hinder the photon propagation in FSO channel.
The geometric losses occur due to the spreading of the beam propagating from the transmitter to the receiver \cite{geo1-1, detail,losses_book}.
It can be calculated as $\left[\frac{d_{r}}{d_{t}+D\text{\ensuremath{L}}}\right]^{2}$, where $d_{r}$
and $d_{t}$ are the diameters of the receiver and transmitter apertures, respectively,
$D$ is the beam divergence and $L$ is the channel length. The atmospheric
attenuation is described by the Beer-Lambert's law $\tau=\text{exp\ensuremath{\left(-\alpha\text{\ensuremath{L}}\right)} dB/km}$, where
$\alpha$  is the atmospheric attenuation coefficient that include the absorption and scattering of the atmospheric medium \cite{losses_book}.
Thus, the total atmospheric loss can be expressed as
\begin{equation}
T=\left[\frac{d_{r}}{d_{t}+D\text{\ensuremath{L}}}\right]^{2}\text{\,exp}\left(-\alpha\text{\ensuremath{L}}\right)\label{eq:2}.
\end{equation}
Table\ref{table1} shows the total channel losses that one can expect from geometrical and atmospheric losses considered for different channel length scales: lab-scale, outside-lab and larger-scale distances. As evident, for large channel lengths, the loss is mainly dominated by the atmospheric losses ($\alpha$) that exponentially increases with \ensuremath{L}. While for smaller channel lengths (lab scale), the channel loss is mainly dominated by the geometrical losses.
\begin{table}[h]
\caption{Typical values of the total channel losses for different channel lengths.}
\vspace{0.3cm}
\label{table1}
\begin{tabular}{|c|c|c|c|c|c|}
\hline  
$L$(m)  & $d_{t}$(mm) & $d_{r}$(mm) & $D$(mrad) & $\alpha$(dB/km) & Channel loss (dB)\tabularnewline
\hline 
\hline 
10 (lab-scale) & 10 & 10 & $0.025$ & $0.1$ & 0.02\tabularnewline
\hline
500 (outside-lab) & 10 & 12 & $0.025$ & $0.1$ & 5.68\tabularnewline
\hline 
30,000 (larger-scale) & 10 & 100 & $0.025$ & $0.1$ & 30.64\tabularnewline
\hline 
\end{tabular}
\par\end{table}
  
\section{Single-photon based QKD protocols (BB84 and six-state)\label{sec: BB84_and_B92}}
QKD protocols, 
based on single-photons, are routinely implemented using attenuated pulsed lasers due to the challenge of obtaining sources of true single-photons. Such attenuated laser sources  are prone to information leakage through photon number splitting (PNS) attack due to multiphoton pulse generation. 
A need for true single photon sources have led to the research efforts  in color centers \cite{Aharonovich}, quantum dots \cite{Pelton}, atoms \cite{McKeever}, trapped  ions in a cavity \cite{Blatt}, etc. Another popular technique for single-photon sources exploits the second-order susceptibility $(\chi^{(2)})$ of a nonlinear material through spontaneous parametric down-conversion (SPDC) in which an intense laser interacts with a nonlinear material to generate two down-converted photons (idler/signal). Conditioned on the detection of an idler photon, the signal photon can be used as a resource of heralded single photon source.
Such heralded sources can be  a promising candidate for implementing the single photon based QKD protocols such as BB84 \cite{BB84} and six-state \cite{six-state}.

BB84 is a four non-orthogonal state-based protocol in which Alice (sender) prepares a string of single photons in one of the four polarization states  $ |\uparrow\rangle,\,|\downarrow\rangle,\,|\nearrow\rangle\text{\,and}\,|\nwarrow\rangle$ and send it to Bob (receiver) who randomly performs the measurement in the rectilinear $\{|\uparrow\rangle, |\downarrow\rangle\}$  or  diagonal $\{|\nearrow\rangle, |\nwarrow\rangle\}$ bases and use the photons for key generation measured in the same basis.
An extended version of BB84 (six-state protocol) has been proposed with more tolerance to noise that enhances the security compared to BB84 \cite{six-state}. In this protocol, six states $ |\uparrow\rangle,\,|\downarrow\rangle,\,|\nearrow\rangle, \,|\nwarrow\rangle, |\circlearrowright\rangle,\,\text{and}\,|\circlearrowleft\rangle$
in three bases \footnote {where $|\uparrow\rangle(|\downarrow\rangle)=|H\rangle(|V\rangle), |\nearrow\rangle(|\nwarrow\rangle)=\frac{|H\rangle+|V\rangle}{\sqrt{2}}(\frac{|H\rangle-|V\rangle}{\sqrt{2}}),\,|\circlearrowright\rangle(|\circlearrowleft\rangle)=\frac{|H\rangle+ i|V\rangle}{\sqrt{2}}(\frac{|H\rangle- i|V\rangle}{\sqrt{2}})$.} (rectilinear, diagonal and circular) is used. The extra choice of basis creates an obstacle on Eve\textquotesingle s measurement path and produces higher error rate. Consequently, Alice and Bob can easily detect the Eve\textquotesingle s presence. In both the protocols, Alice and Bob keep the photons which are measured in the same basis $(\frac{1}{2}$ and $\frac{1}{3}$ probability for BB84 and six-state protocol, respectively \cite{Rev2020}).
\begin{figure}[t]
\begin{centering}
\includegraphics[width=0.7\linewidth]{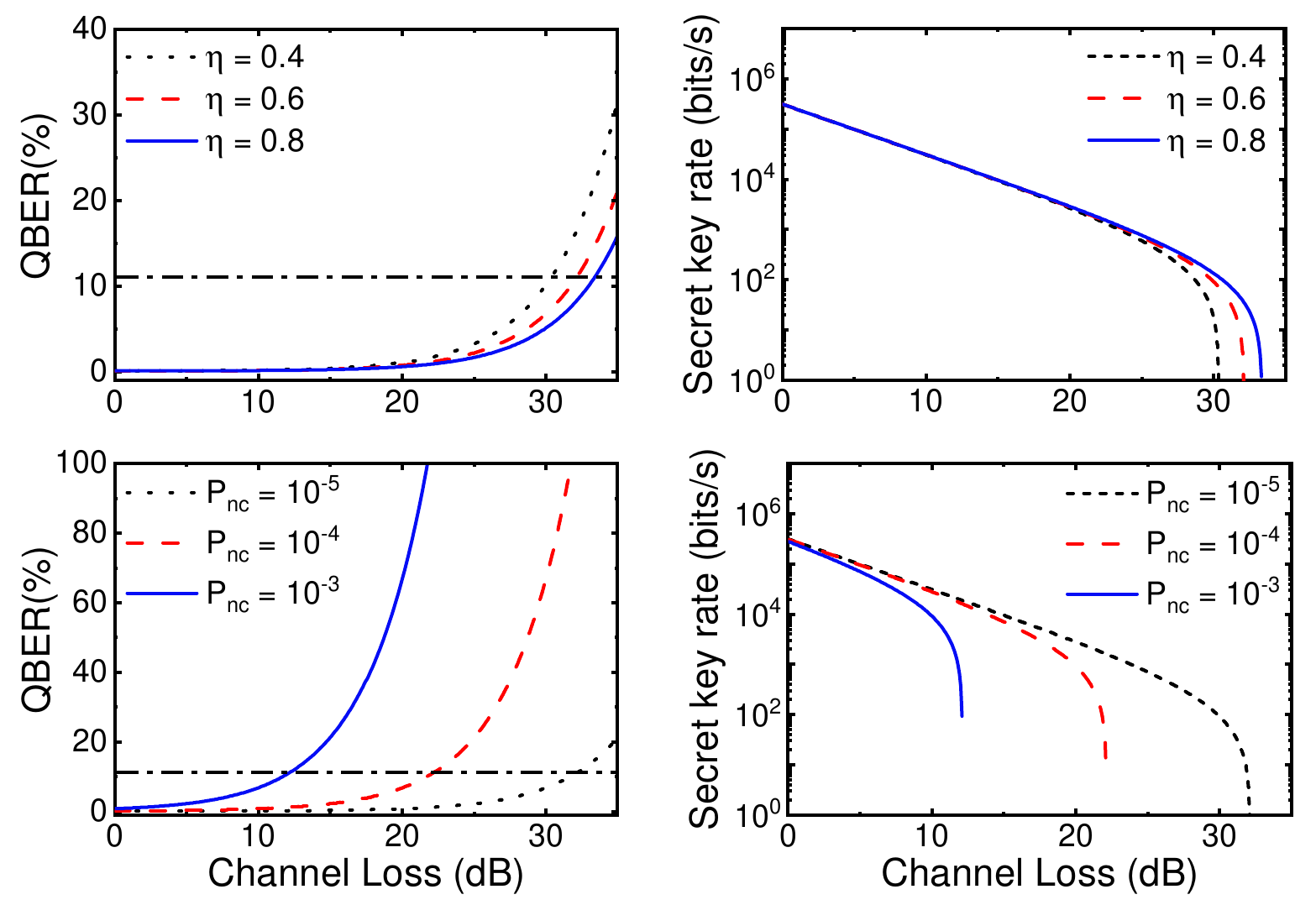}
\par\end{centering}
\caption{\label{fig:BB84_QBER_SKR}Variation of QBER and SKR for BB84 protocol as a function of channel loss ($T$) for different detector efficiency ($\eta$) and noise count probability ($P_{nc}$).}
\end{figure}
For both the protocols, the error in the shifted key is calculated by measuring the QBER [Eq.\ref {eq:1}]. There are mainly two contributions to $N_{error}$ ($P_{opt}$ and $P_{nc}$) for single-photon based protocols. Thus, the total QBER can be calculated as \cite{Gisin}
\begin{equation}
Q = P_{opt} + \beta\frac{P_{nc} n}{T \eta q \mu},
\label{qber}
\end{equation}
where $\beta = \frac{1}{2}$ for BB84 protocol and $\frac{2}{3}$ for the six-state protocol, $P_{opt}$ is the probability of incorrect detections of the photons due to imperfect interference or polarization contrast, $P_{nc}$ is the probability of overall noise counts that include the detector dark counts and the background counts, 
$q$ (1 or 0.5) is used to correct the non-interfering path combinations, $n$ is the number of detectors, $\eta$ is the detector efficiency, $T$ is the channel transmittance, and $\mu$ is the mean photon number ($\mu = 1$ for single photon sources).  
The secret key rate for BB84 and six-state protocol is calculated as \cite{Rev2020}  

\begin{equation}
S_{BB84}=\frac{1}{2}\nu_{S}T\left[1+2Q\,\text{lo\ensuremath{g_{2}}\ensuremath{Q}+2\ensuremath{(1-Q)}\,\text{lo\ensuremath{g_{2}}}\ensuremath{(1-Q)}}\right]\label{eq:SKRBB},
\end{equation}

\begin{equation}
S_{six-state}=\frac{1}{3} \nu_{S} T\left[1+\frac{3Q}{2}\,\text{lo\ensuremath{g_{2}}\ensuremath{\ensuremath{\frac{Q}{2}}+}\ensuremath{\left(1-\frac{3Q}{2}\right)}\,\text{lo\ensuremath{g_{2}}}\ensuremath{\left(1-\frac{3Q}{2}\right)}}\right]\label{eq:sixSKR}.
\end{equation}

where $\nu_{S}$ is the heralded single photon counts at the sender's side. For the present study, we have considered a type-0 SPDC source with a brightness (photon-pairs per unit mW pump power) of $\nu_{S} = 0.64\times10^6$ cps/mW from Ref. \cite{Steinlechner2012}. 
\begin{figure}[t]
\begin{centering}
\includegraphics[width=0.7\linewidth]{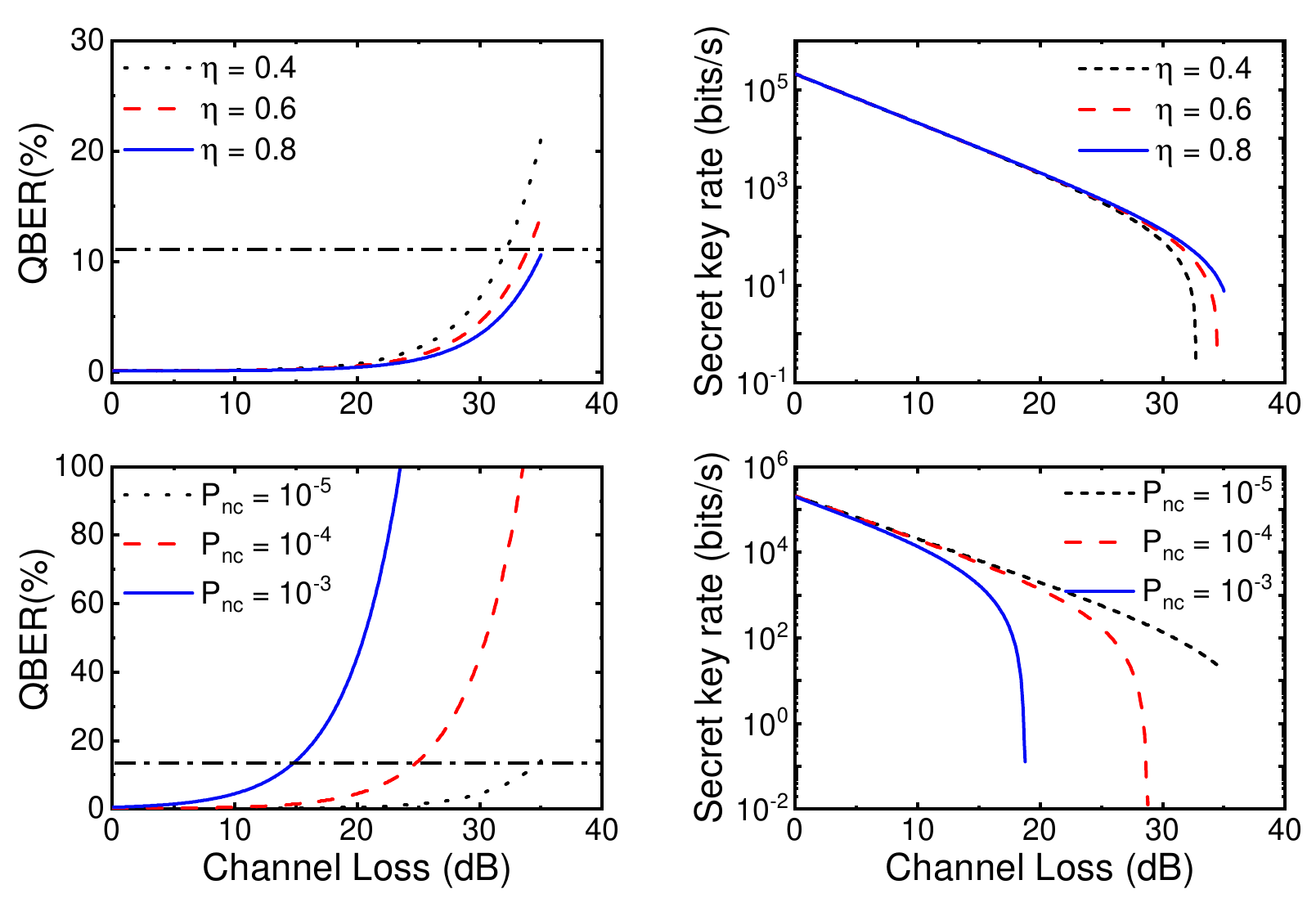}
\par\end{centering}
\caption{\label{fig:B92}Variation of QBER and SKR  for six-state
protocol  as a function of channel loss $(T)$ for different detector efficiency ($\eta$) and noise count probability ($P_{nc}$). For top row graphs, $P_{nc} = 10^{-5}$ and for bottom row graphs, $\eta = 0.6$.}
\end{figure}

Figures \ref{fig:BB84_QBER_SKR} and \ref{fig:B92} show the calculated QBER and SKR for BB84 and six-state protocols, as a function of the channel loss at different detector efficiencies ($\eta=0.4,\,0.6,\,0.8$) and noise count probabilities ($P_{nc}=10^{-5},\,10^{-4},\,10^{-3}$) for $q=0.5$, $\mu=1$, $P_{opt}=0.001,$ $\nu_{S}=0.64\times10^6 \text{ cps}$ , $n=4$. In these figures, the black dashed-dotted line shows the threshold value of QBER for the respective protocols. A maximum threshold of $11\%$ and $12.6\%$ ($\eta=0.4$, black dotted line in Figures \ref{fig:BB84_QBER_SKR} and \ref{fig:B92}) for BB84 and six-state protocol yield a noise tolerance of 33dB and 36dB, respectively. The noise tolerance increases with an increase (decrease) of the  detector efficiency (noise count probability). Moreover, the SKR is high when channel losses are low and decreases sharply to the threshold limit of 33dB and 36dB, respectively, corresponding to the threshold value of QBER. 
We infer that detectors with high efficiency, that affect the overall noise counts, are required to tolerate high channel losses or longer channel lengths.

\begin{figure}[h]
\begin{centering}
\includegraphics[width=0.5\linewidth]{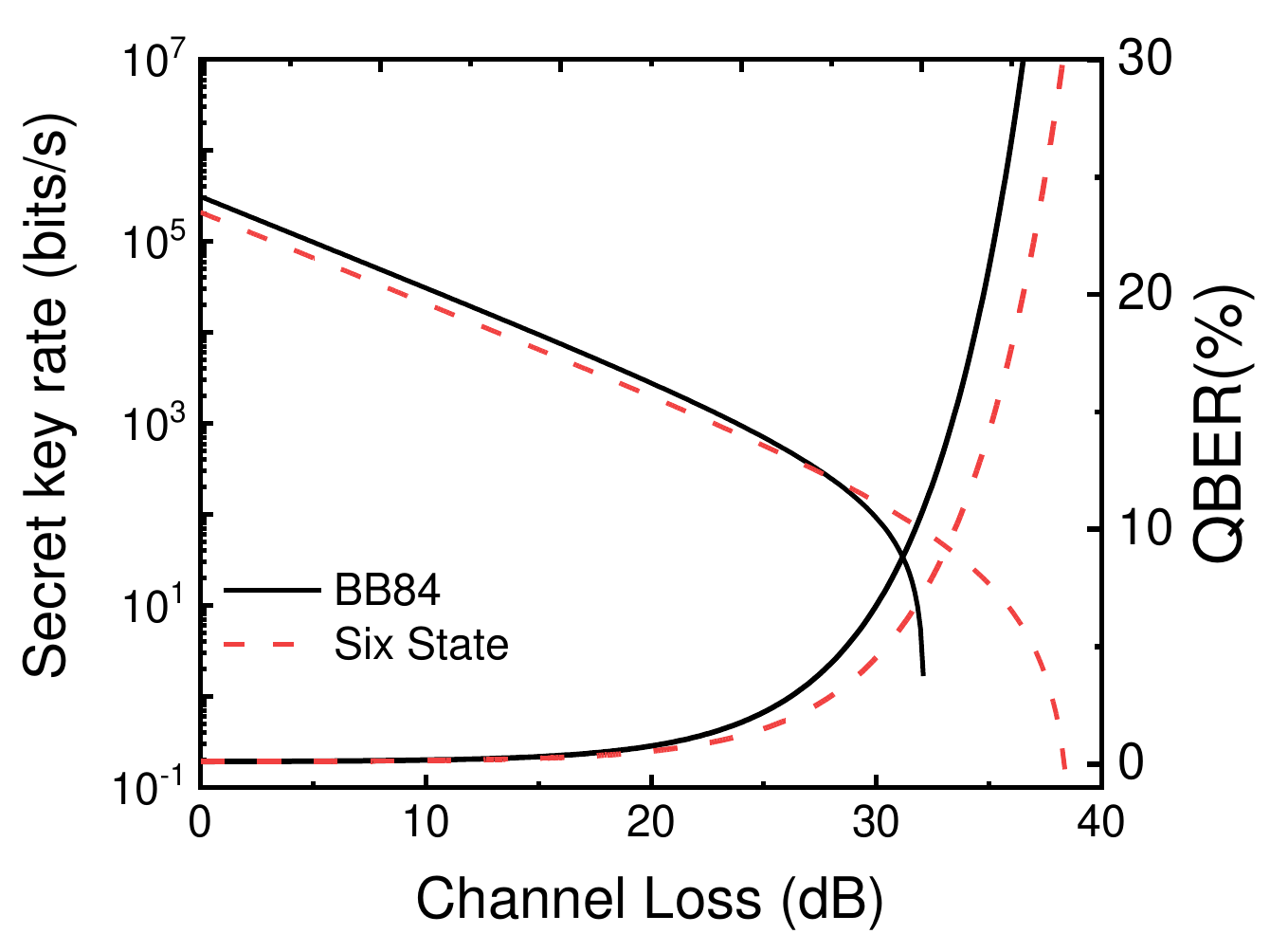}
\par\end{centering}
\caption{\label{fig:com}Comparison between the BB84 and six-state protocol
 with respect to channel loss for $\eta = 0.6$ and $P_{nc} = 10^{-5}$. }
\end{figure}

Figure \ref{fig:com} shows the calculated SKR of BB84 and six-state protocol at different channel losses for $\eta=0.6$  and $P_{nc}=10^{-5}$. Since, the six-state protocol utilizes three MUBs, more information about the Eve's presence can be obtained, resulting in a higher bit error rate threshold and higher noise tolerance.
\section{Entanglement-based QKD protocols \label{sec:Ekert91-and-BBM92}}

Artur Ekert proposed the first entanglement-based QKD protocol \cite{Ekert} by exploiting the maximally entangled states that violate the Clauser-Horne-Shimony-Holt (CHSH)
inequality \cite{Bell}. It utilizes the three randomly selected bases to measure
the polarization of the entangled-photon. The extra basis is required to perform the Bell's inequality
test that directly
detects the presence of an eavesdropper without revealing the key information. Nonlinear optical techniques like, SPDC and spontaneous four-wave mixing (SFWM) have been capitalized to generate the polarization entangled-photon. Such entangled-states intrinsically increase the security of the shared information, govern by the inherent quantum nature of the source. The degree of violation of a Bell inequality is used to quantify the quality of entanglement between the photon-pairs, for which Bell parameter $(S_{CHSH})$  is calculated as
\begin{equation}
S_{CHSH}=|E\left(\theta_{A}^{1},\,\theta_{B}^{1}\right)+E\left(\theta_{A}^{1},\,\theta_{B}^{3}\right)-E\left(\theta_{A}^{3},\,\theta_{B}^{1}\right)+E\left(\theta_{A}^{3},\,\theta_{B}^{3}\right)|,\label{eq:Bell}
\end{equation}

where $E\left(\theta_{A}^{i},\,\theta_{B}^{j}\right)$ is the correlation
coefficient at two different orientation of the analyzers of Alice and Bob corresponding
to different chosen bases, $i$ and $j$, respectively.

Here are the different cases of the calculated $S$ parameter:
\begin{itemize}
\item $|S_{CHSH}|\leq2$ implies an extreme case of destruction of entanglement signifying the classical nature of the source with  no chance of a key generation.
\item $|S_{CHSH}|=2\sqrt{2}$ represents a perfectly entangled state (maximum value of the violation of the Bell's inequality). This is an ideal scenario.
\item For $2<|S_{CHSH}|<2\sqrt{2}$ implies a real situation with Eve's presence and/or noise detection. The sifted key may not be discarded and can be used to generate
a secret key after classical post-processing  algorithms, used for error correction
and privacy amplification.
\end{itemize}

We consider a SPDC source that generates entangled photon-pairs and sends them through the turbulent medium (atmosphere) to the receiver stations A(Alice) and B (Bob). 
One of the maximally entangled polarization state can be expressed as:

\begin{equation}
|\psi\rangle=\frac{1}{\sqrt{2}}\left(|H\rangle_{A}|V\rangle_{B}+e^{i\varphi}|V\rangle_{A}|H\rangle_{B}\right), \label{eq:Bell-1}
\end{equation}
where, $H$ and $V$ are the polarizations of the photon-pair and $\varphi$ is the relative phase.
The randomly chosen orientation of the analyzer angles at A and B are $\left(\theta_{A}^{1},\,\theta_{A}^{2},\,\theta_{A}^{3}\right.$
$\left.=0,\,\frac{\pi}{8},\,\frac{\pi}{4}\right)$ and $\left(\theta_{B}^{1},\,\theta_{B}^{2},\,\theta_{B}^{3}=-\frac{\pi}{8},\,0,\,\frac{\pi}{8}\right)$, respectively. For $\varphi =\pi$ and $\left(\theta_{A}^{1},\,\theta_{A}^{3},\,\theta_{B}^{1},\,\theta_{B}^{3}\,\right)$ =
$\left(0,\,\frac{\pi}{4},\,-\frac{\pi}{8},\,\frac{\pi}{8}\right),$
a maximal value $(2\sqrt{2})$ of $S_{CHSH}$ is reached for the $|\psi\rangle$
state, for which the correlation coefficient can be calculated as \cite{Bell_parameter} 
\[
E\left(\theta_{A},\,\theta_{B}\right)=N\left[-\text{cos}2\theta_{A}\text{cos}2\theta_{B}+\text{cos}\varphi\,\text{sin}2\theta_{A}\text{sin}2\theta_{B}\right],
\]
where $N$ is a constant that depends on the noise counts and atmospheric losses, leading to a reduction in $S_{CHSH}$ and defined as \cite{Bell_parameter}                              
\[
N=\frac{p_{s}\eta_t^{2}}{p_{s}\left[\eta_t+2P_{nc}\left(1-\eta_t\right)\right]^{2}+2p_{1}P_{nc}\left[\eta_t+2P_{nc}\left(1-\eta_t\right)\right]+4p_{0}P_{nc}^{2}}.
\]
where, $p_{s}=T_{A}\cdot T_{B}$   is the  Bell state probability with $T_{A\left(B\right)}$ (Eq. \ref{eq:2}) being the tranmission cofficient of the receiver stations A(B) \footnote{We have considered Alice's and Bob's stations to be exposed to equal losses ($T_{A}=T_{B}$).}, $p_{1}=p_{H_{A}}+p_{V_{A}}+p_{H_{B}}+p_{V_{B}}$ is the sum of the probability of a single-photon state calculated as: $p_{H_{A}}=\frac{1}{2}\left(T_{A}\left(1-T_{B}\right)\right)$, $p_{H_{B}}=\frac{1}{2}\left(T_{B}\left(1-T_{A}\right)\right)$, $p_{V_{A}}=p_{H_{A}}$, $ p_{V_{B}}=p_{H_{B}}$ ($H$ and $V$ are the horizontal and vertical modes, respectively). Here $\eta_t$  is the total detection efficiency which is the product of detector efficiency ($\eta$) and photon collecting efficiency ($\eta_c = 0.6$).

\begin{figure}
\begin{centering}
\includegraphics[width = 0.8\linewidth]{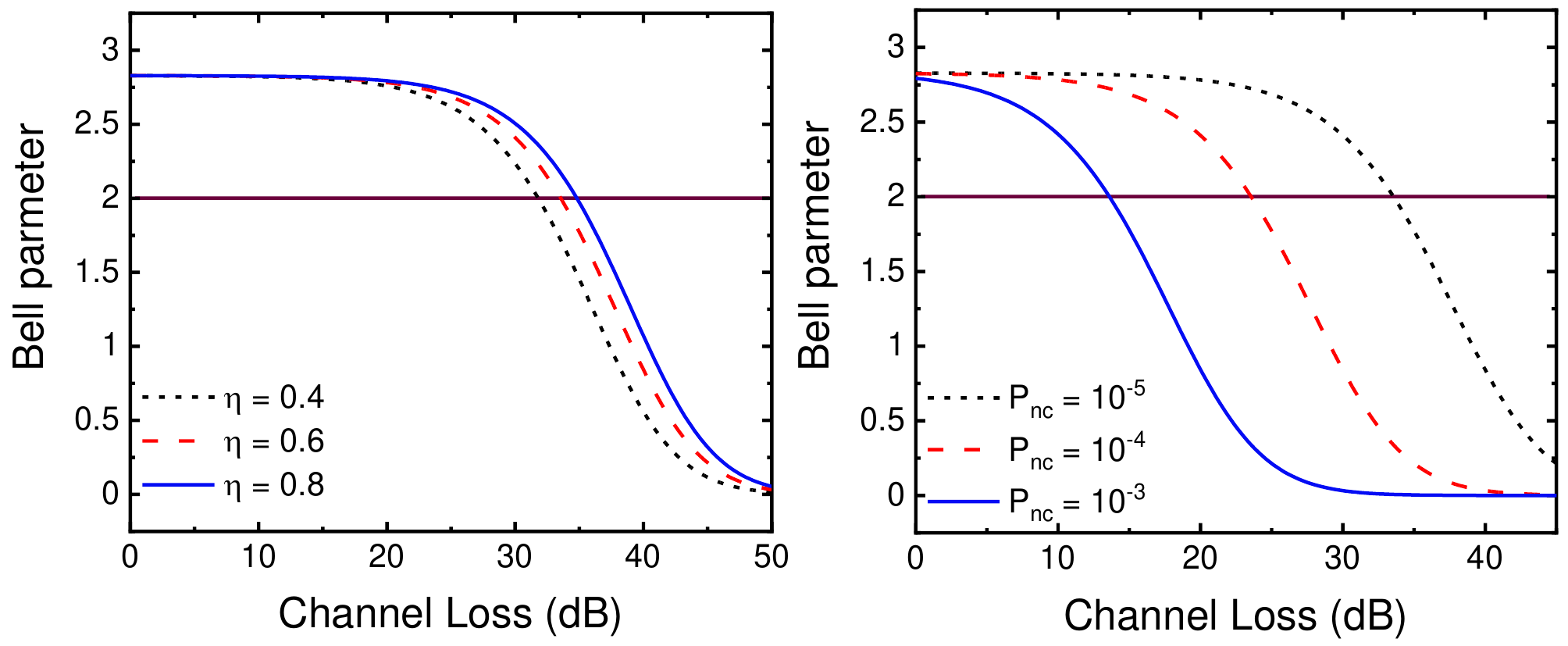}
\par\end{centering}
\caption{\label{fig:bell} Effect of the channel losses on the entanglement quality for different detection efficiencies and noise counts.} 
\end{figure}

The variation of the Bell parameter as
a function of channel loss with different detection efficiencies ($\eta$) and probability of noise counts $\left(P_{nc}\right)$ is shown in Figure \ref{fig:bell}. It is evident that entanglement can survive and tolerate higher losses
for lower noise counts (black dashed line). The purple solid line
shows the threshold of Bell's inequality ($S_{CHSH}=2$) signifying a witness of the entanglement or non-locality (quantum
phenomenon). 
Although an implementation of the Ekert's (E91) protocol is partly complicated as it requires the Bell's inequality test to detect  Eve's presence,
however it has the quality of utmost and unconditional security (more secure even for devices that are not trusted)
and can be used in a special case where other protocols (e.g., BBM92,
BB84)  fail to perform. The E91 protocol is a fully device-independent QKD
(DIQKD) protocol, which facilitates the unconditional security without any trusting the QKD device  \cite{DI-QKD}. 

\begin{figure}[h]
\centering{}\includegraphics[width = 0.7\linewidth]{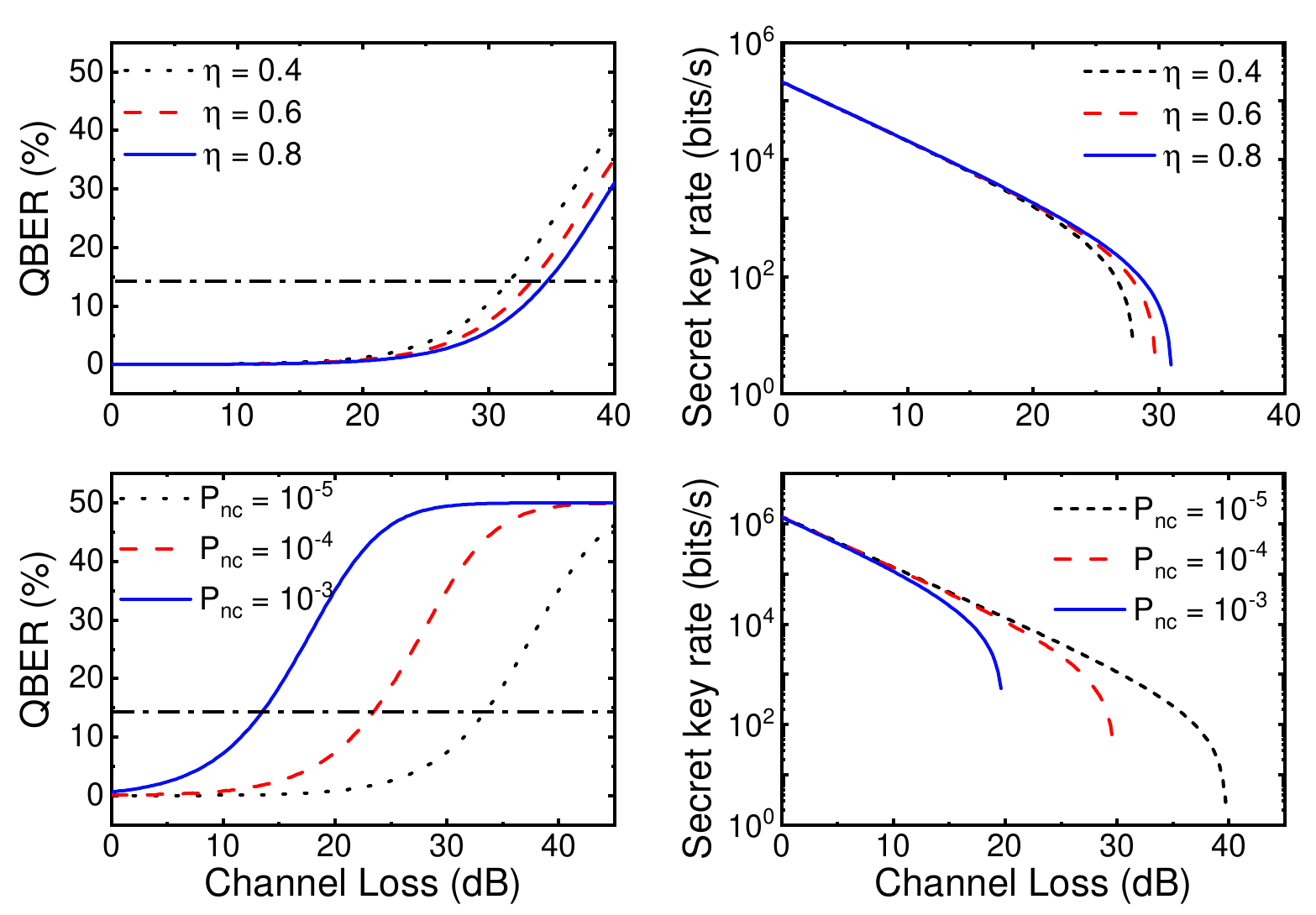}\caption{\label{fig:Variation-of-}Plots of the QBER  and SKR in terms of  channel loss with varying $\eta$ and $P_{nc}$ for E91 protocol. For top row graphs, $P_{nc} = 10^{-5}$ and for bottom row graphs, $\eta = 0.6$.}
\end{figure}

In 2007 \cite{Acin}, Acin et al. derived a formula for unconditional
security bound $I_{E}=h\left(\frac{1+\sqrt{S^{2}/4-1}}{2}\right)$
and the relation $S_{CHSH}=2\sqrt{2}\left(1-2Q\right)$ for the E91 protocol. 
It is
clear that when $Q\cong14.6\%$ then  $S_{CHSH}=2,$ and therefore no secure key can be
generated. In Figure \ref{fig:Variation-of-}, we have plotted the QBER and SKR as a function of
channel loss for different detection efficiences and probabilities noise counts 
using 

\begin{equation}
S_{E91}=\frac{1}{3}\nu_{s}T\left[1-h(Q)-h\left(\frac{1+\sqrt{S_{CHSH}^{2}/4-1}}{2}\right)\right].\label{eq:91}
\end{equation}

The necessary condition (Bell's inequality violation) or the need of an extra
basis to calculate the amount of information leaked by eavesdropper
was removed by Bennett, Brassard, and Mermin in their 1992 protocol (BBM92) \cite{bbm_92}.
In this entanglement-based BBM92 protocol, 
Alice and Bob use two mutually
unbiased bases (rectilinear or diagonal) to randomly perform the measurement
on the entangled photons (entanglement version of the BB84 protocol).
The classical error correction and privacy amplification part is similar to the BB84 protocol. 
In this protocol, a central SPDC source generates entangled photon pairs, one of which is sent to Alice while the other photon is sent to Bob. Alice and Bob randomly choose the basis to perform the measurement on the incoming
photons. In an ideal scenario (no eavesdropping), if Alice and Bob
choose the same basis, then their measurement outcome will always be same or completely correlated  
as the two photons of an Einstein\textendash Podolsky\textendash Rosen
(EPR) pair are correlated. Thus a symmetric key (sifted key) is generated
(which is discarded when they measure in the different bases). Then,
they perform classical error correction and privacy amplification
to estimate the QBER. Any attempt of an eavesdropper to intervene (on the source or on
the photons) will destroy the entanglement and introduce an error
in the sifted key. This is the witness of entanglement in BBM92 protocol,
while Eve's information is bounded by evaluating Bell's inequality
whose violation (non-locality) is the witness of entanglement in Ekert91 protocol. 

In an entanglement-based QKD protocol, imperfections in the entangled photon
pair sources are characterized by the two photon interference visibilites based on the polarization correlations
$V_{HV}$ and $V_{\pm45}$ in the $HV$ and $\pm$ bases, respectively. Intrinsic QBER of a QKD system is
calculated as $q_{i}=\frac{1-V_{tot}}{2}$, where $V_{tot}=\frac{V_{HV}+V_{\pm45}}{2}$ 
 which can be directly calculated  by performing two-photon interference measurements
\cite{BBM4}.

%

We consider two situations in the  BBM92 protocol:

(1) When the source at Alice's side 

(2) When the source placed in the middle.

In both cases, the raw key rate is half of the detected coincidence rate,
\begin{figure}[h]
\begin{centering}
\includegraphics[width =0.7\linewidth]{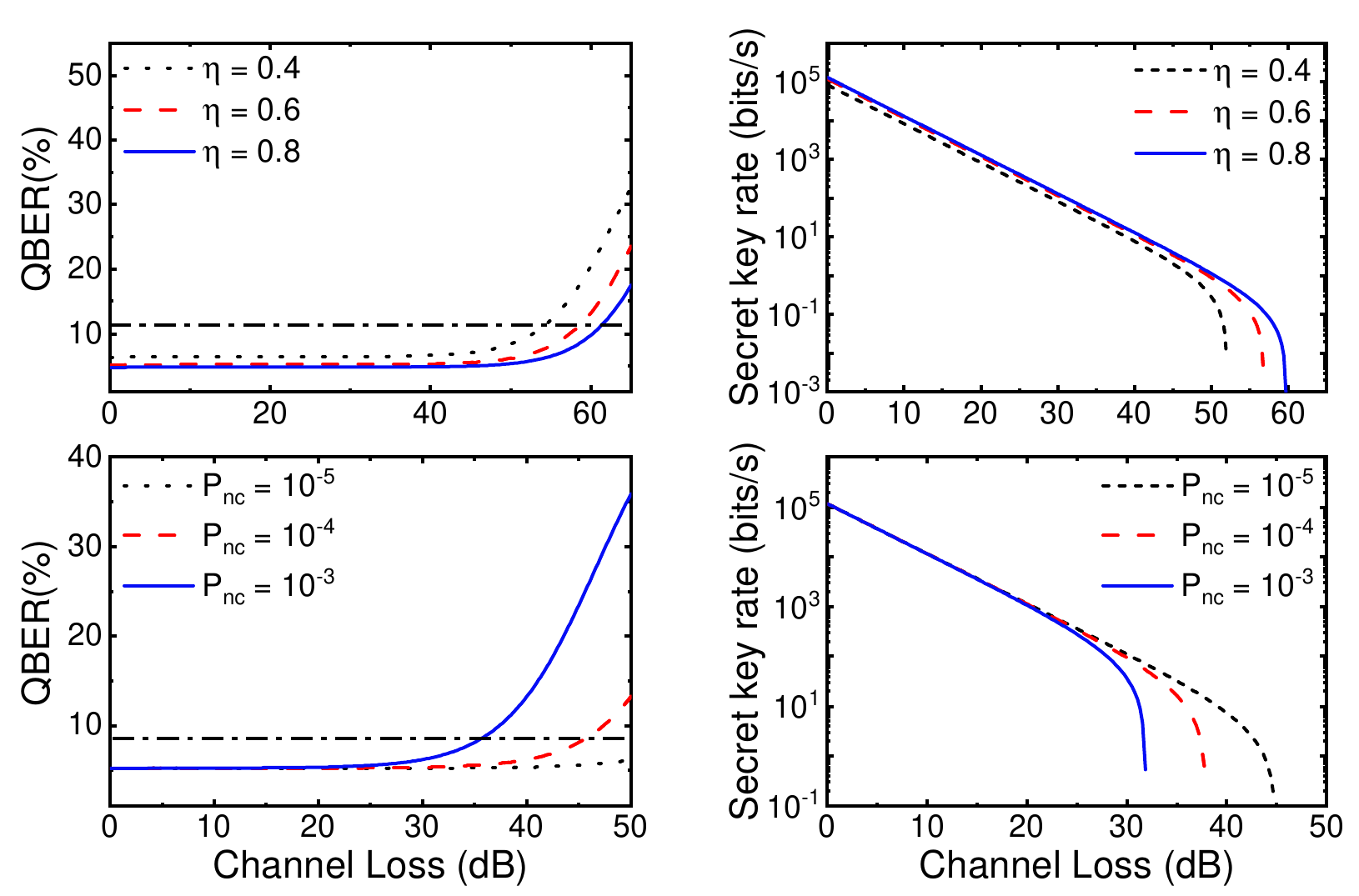}
\par\end{centering}
\caption{\label{fig:BBM92_Alice's_side}Variation of QBER and SKR with respect
to channel loss for different values of $\eta$ and $P_{nc}$,  when source
at Alice's side (BBM92). For top row graphs, $P_{nc} = 10^{-5}$ and for bottom row graphs, $\eta = 0.6$.}
\end{figure}
\begin{equation}
r_{sig}=\frac{1}{2}r_{c}T,\label{eq:4}
\end{equation}

\begin{figure}
\begin{centering}
\includegraphics[width = 0.7\linewidth]{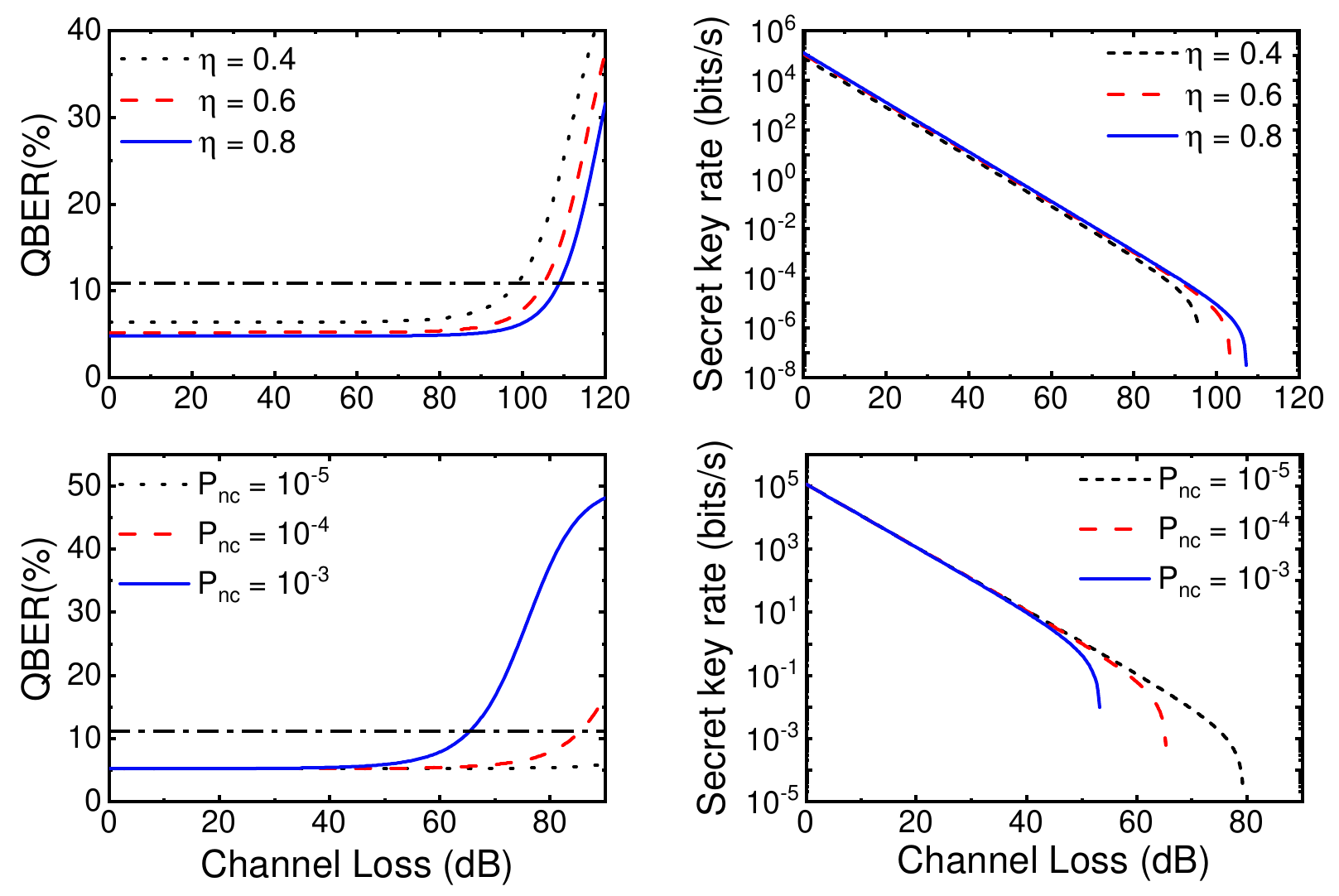}
\par\end{centering}
\caption{\label{fig:BBM92@-source-in}QBER and SKR  with varying $\eta$ and $P_{nc}$ in terms of 
 channel loss, when source placed in the middle (BBM92). For top row graphs, $P_{nc} = 10^{-5}$ and for bottom row graphs, $\eta = 0.6$.}
\end{figure}
 where $r_{c}$ is the coincidence rate corresponding to  single event rates $r_{1}$
(Alice's detector) and $r_{2}$ (Bob's detector) which include detector
efficiencies and $T$ is the transmission of the entire quantum channel.
The accidental coincidence rate where only one detector is exposed to
the background events is given by \cite{Daylight} ,

\begin{equation}
r_{a(single)}=\frac{1}{2}\left(r_{1}-Tr_{c}\right)\left(r_{bg}+T\left(r_{2}-r_{c}\right)\right)\tau_{c},\label{eq: 5}
\end{equation}

When the source is in the middle, both detectors are exposed to the
background events in which case the accidental coincidence rate is expressed
as:

\begin{equation}
r_{a(both)}=\frac{1}{2}\left(r_{bg}+T\left(r_{1}-r_{c}\right)\right)\left(r_{bg}+T\left(r_{2}-r_{c}\right)\right)\tau_{c},\label{eq:6}
\end{equation}

where $r_{bg}$ is an external background event rate calculated as $r_{bg} = P_{nc}r_1(r_2)$, $\tau_{c}$
is a coincidence time interval. The total QBER
is \cite{Daylight},

\begin{equation}
Q=\frac{1}{r_{sig}+r_{a}}\left(q_{i}r_{sig}+\frac{1}{2}r_{a}\right).\label{eq:-7}
\end{equation}

We have adapted the value of the parameters from an experimental
study that considers \textcolor{black}{\cite{Steinlechner2012}}:
 $\nu_{S} =$ $r_1 =$  $r_2 = 0.64 \times 10^6$, $r_c = \eta^2 \eta_c^2r_1$;
where $\eta_c$ is the photon collection efficiency into the fiber and $\eta$ is the detector efficiency, and $\tau_{c}=2 \text{ ns}$, $q_{i}=0.043$ \cite{Daylight}. 

When the source is in the middle both arms are exposed to  detector error, background counts and other
 losses. Hence there impact is doubled in comparison to the situation
when the source is at Alice's or Bob's side (only one arm exposed to the
losses and errors), therefore QBER is always higher in the second
case of BBM92. 
Figures \ref{fig:BBM92_Alice's_side} and \ref{fig:BBM92@-source-in} show the effect of free-space losses on QBER and SKR for different detector efficiencies and noise count probabilities.
\begin{figure}[h]
\begin{centering}
\includegraphics[width = 0.5\linewidth]{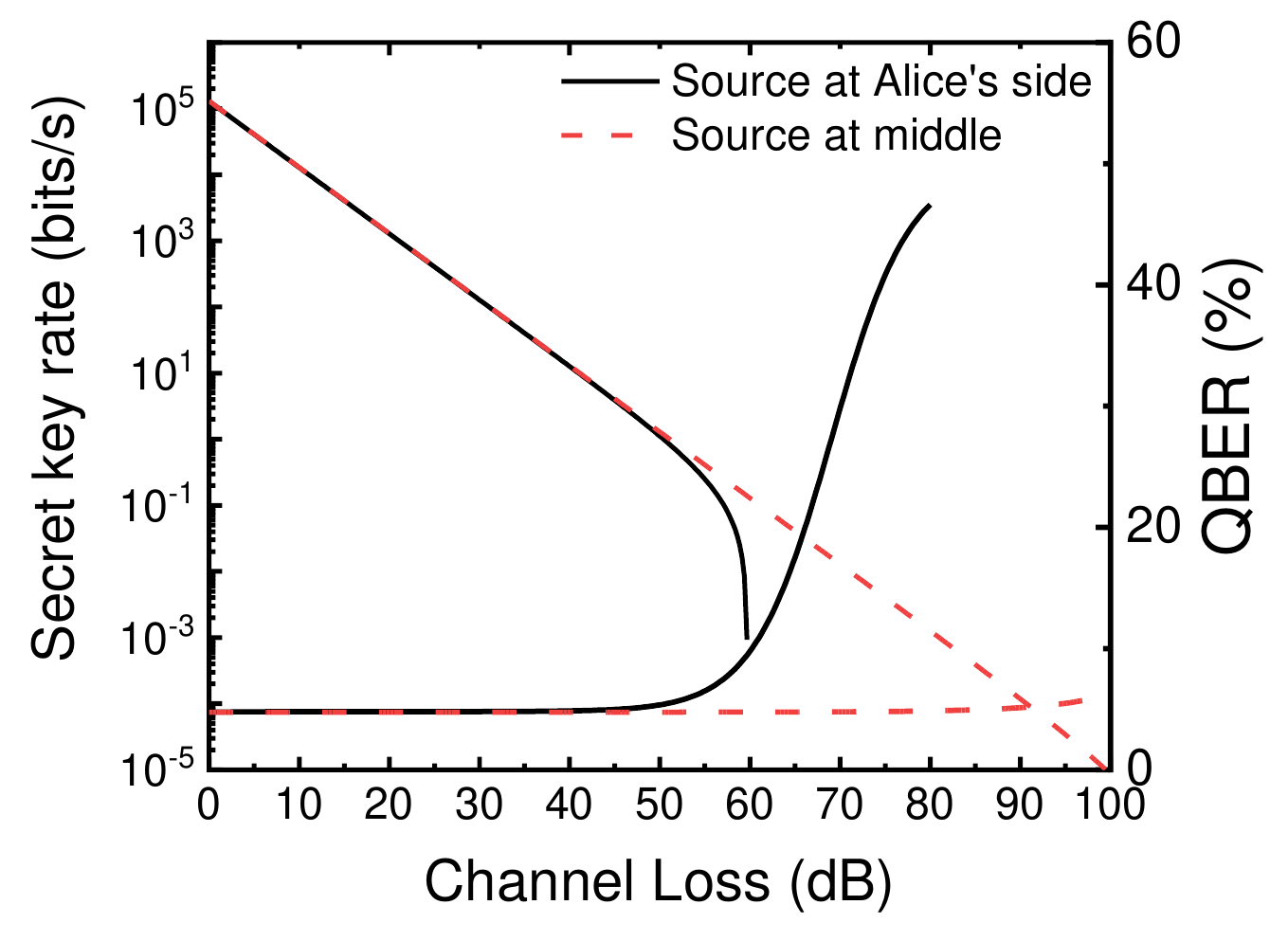}
\par\end{centering}
\caption{\label{fig:Comparison-BBM92}Comparison of BBM92 QKD
protocol as a function of channel losses when source is Alice \textquotesingle s side and at the middle for $\eta = 0.6$ and $ P_{nc} = 10^{-5}$.}
\end{figure}
The entanglement-based QKD is a basis-independent QKD because the state emitted
from the EPR source is independent of the measurement bases in Alice's and
Bob's side \cite{Ma}. The unconditional security in these cases are established
by Koashi and Preskill \cite{koashi} and improved by Ma, Fung, and
Lo \cite{Ma}. The secret key generation rate for the BBM92 protocol at
the  QBER threshold value of 11\% is,

\begin{equation}
S_{BBM92}=\frac{1}{2}\nu_{S}T\left[1-f\left(Q\right)h_{2}\left(Q\right)-h_{2}\left(Q\right)\right].\label{eq:8}
\end{equation}
here, $f\left(Q\right)$ is the bidirection error correction efficiency
as a function of error rate. The values of $f\left(Q\right)$ for different
error rates may be found  in Ref. \cite{fe}. 

Figure \ref{fig:Comparison-BBM92} shows the QBER and SKR comparing the two above situations of the BBM92 protocol. We see that
when the source is in the middle, the protocol tolerates higher channel losses, almost double compared  to the situation when the source is at Alice's side.

\section{Conclusion\label{sec:Conclusion}}

We have theoretically studied and compared four different QKD protocols based on single-photon (BB84 and six-state) and entanglement-photons (Ekert91 and BBM92) by evaluating the QBER and SKR for free-space losses and device imperfections. 
The role of detector efficiency and noise counts for different channel losses is studied and shown in Figures \ref{fig:BB84_QBER_SKR}, \ref{fig:B92},  \ref{fig:bell}-\ref{fig:BBM92@-source-in}.
It is shown that an increment in the channel loss 
leads to a higher QBER resulting in a lower SKR for both single-photon and entangled-photon based QKD protocols.  The detector efficiency greatly affects the QBER
and SKR for larger channel loss (channel length) due to the exponential rise of
channel losses compared to lower channel lengths (especially in lab-scale implementations). The numerically calculated values of the QBER and SKR are considered for QKD protocols at different length scales: lab-scale, outside-lab and larger-scale distances is shown in Table \ref{table2}.

\begin{table}[h]
\caption{The expected QBER and SKR for different QKD protocols at different channel lengths for $\eta = 60\%$ and $P_{nc} = 10^{-5}$.}
\vspace{0.5cm}
\label{table2}
\begin{tabular}{|c|c|c|c|c|c|c|c|c|c|}
\hline 

\multicolumn{1}{|c|}{Channel} &\multicolumn{2}{c|}{BB84}&\multicolumn{2}{c|}{Six-state} &\multicolumn{2}{c|}{BBM92}&\multicolumn{2}{c|}{E91}\\
\cline{2-9}
\multicolumn{1}{|c|}{Length} &\footnotesize{}QBER &\footnotesize{}SKR&\footnotesize{}QBER &\footnotesize{}SKR&\footnotesize{}QBER &\footnotesize{}SKR&\footnotesize{}QBER &\footnotesize{}SKR\\
\multicolumn{1}{|c|}{}&(\footnotesize{}\%) & \footnotesize{}(bits/sec)&\footnotesize{}(\%) & \footnotesize{}(bits/sec)&\footnotesize{}(\%) &\footnotesize{}(bits/sec)&\footnotesize{}(\%) & \footnotesize{}(bits/sec)\\
\hline
\footnotesize{}10 m &\footnotesize{}0.107&\footnotesize{}$3.11\times 10^5$&\footnotesize{}0.105&\footnotesize{}$2.09\times 10^5$&\footnotesize{}5.18&\footnotesize{}$1.16\times 10^5$&\footnotesize{}0.006&\footnotesize{}$2.1\times10^5$\\
\footnotesize{}500 m  &\footnotesize{}0.125&\footnotesize{}$0.84\times 10^5$&\footnotesize{}$0.12$&\footnotesize{}$0.57\times 10^5$&\footnotesize{}5.22&\footnotesize{}$0.31\times 10^5$&\footnotesize{}0.007&\footnotesize{}$1.8\times 10^5$\\
\footnotesize{}30 km  &\footnotesize{}7.6&\footnotesize{}86&\footnotesize{}5.21&\footnotesize{}$132$&\footnotesize{}5.24&\footnotesize{}$106$&\footnotesize{}7.17&\footnotesize{}4.42\\
\hline
\end{tabular}
\par
\end{table}

Since, the atmospheric losses are inevitable  and cannot be controlled,  to obtain a low QBER, near-to-perfect devices are desirable that have high efficiency and minimal losses.
In this study, we have considered  practical values of different
parameters (detector's efficiency, background counts, coincidence rates, 
diameter of the receiver and transmitter apertures, beam divergence
etc.) for different QKD protocols pertaining to practical systems. We have shown that secret key generation is possible even under atmospheric losses within certain ranges of parameter values. Also, a comparative study of the protocols under the single photon/ prepare-and-measure technique as well as the entanglement-based technique (Figures \ref{fig:com} and \ref{fig:Comparison-BBM92}) show that the single (entangled) photon based six-state (BBM92) protocol tolerates higher channel losses compared to BB84 (E91) protocol. Two cases of source (entangled photon) position for  BBM92 protocol are considered and compared (Figure\ref{fig:Comparison-BBM92}), which proves that when the source is placed in the middle it can tolerate higher channel losses (almost double) as compared to the situation when the source is placed at Alice's side. A benefit of such comparative studies is to facilitate researchers with parameters and values from a practical consideration helping them to select high performace based QKD protocol for free-space under considered atmospheric conditions. The present theoretical work
can be utilized  by experimentalists to implement practical QKD
protocols under different conditions and can be extended for longer distances or for satellite based applications.

\textbf{Acknowledgment: }The authors thankfully acknowledge the following
funding agencies: Council of Scientific Industrial Research (CSIR),
India (09/086(1331)/2018-EMR-I) and Defence Research and Development
Organisation, India (DFTM/03/3203/P/07/JATC-P2QP-07/463/D) for a project
grant.

\end{document}